\documentclass[preprint,aps,prb, superscriptaddress,showpacs,nofootinbib,floatfix, showkeys]{revtex4-2}

\usepackage{graphicx}
\usepackage{dcolumn}
\usepackage{bm}
\usepackage{flafter}
\usepackage{lettrine}
\usepackage{amsmath}
\usepackage{amssymb}
\usepackage{color}
\usepackage{epsfig}
\graphicspath{{eps+pdf/}}
\makeatother
\begin{document}

\date{\today}

\title{Lattice dynamics study of (Bi$_{1-x}$Sb$_x$)$_2$Te$_3$ compounds: Comparative compositional study of  force constants and  Grüneisen parameters.}

\author{Baihong Sun}
\affiliation{Department of Materials Science and Engineering, Guangdong Technion-Israel Institute of Technology, Shantou 515063, China}
\affiliation{Department of Materials Science and Engineering, Technion-Israel Institute of Technology, Haifa 3200003, Israel}
\author{Chenxin Wei}
\affiliation{Department of Materials Science and Engineering, Guangdong Technion-Israel Institute of Technology, Shantou 515063, China}
\affiliation{Department of Materials Science and Engineering, Technion-Israel Institute of Technology, Haifa 3200003, Israel}
\author{Elissaios Stavrou}
\email{Correspondence to: Prof. Elissaios Stavrou,  Materials Science  and Engineering Department, Guangdong Technion - Israel Institute of Technology, 241 Daxue Road, Shantou, Guangdong Province, 515063, China.  E-mail:  elissaios.stavrou@gtiit.edu.cn}
\affiliation{Department of Materials Science and Engineering, Guangdong Technion-Israel Institute of Technology, Shantou 515063, China}
\affiliation{Department of Materials Science and Engineering, Technion-Israel Institute of Technology, Haifa 3200003, Israel}
\affiliation{Guangdong Provincial Key Laboratory of Materials and Technologies for Energy Conversion, Guangdong Technion-Israel Institute of Technology, Shantou 515063, China}

\begin{abstract}
We  performed a comprehensive Raman spectroscopy study of  (Bi$_{1-x}$Sb$_x$)$_2$Te$_3$ (BST-x) compounds, with varying  Sb concentration, under ambient  and   high-pressure conditions. At ambient conditions, the frequency of all observed Raman active modes ($E_g^1$, $A_{1g}^1$, $E_g^2$, $A_{1g}^2$ ) is  blue-shifting with increasing Sb concentration, albeit with varying concentration-dependent rate. Based on these results, we calculated  the relative force constants as a function of Sb concentration using  a  harmonic approximation, which shows a  mode-dependent variation. The force constants for the  lower frequency Raman modes ($E_g^1$ and $A_{1g}^1$) remain almost composition independent, while for the higher-frequency ones a profound softening ($E_g^2$) or stiffening ($A_{1g}^2$) was determined.   From the Raman spectroscopy measurements under high-pressure conditions, the corresponding  Grüneisen parameters for the three observed Raman-active modes ($A_{1g}^1$, $E_g^2$, $A_{1g}^2$)  as a function of composition were determined, by combining them with the previously reported Equation of States and the corresponding bulk modulus B$_0$ of the ambient-conditions  $R\overline{3}m$ phase. The average Grüneisen parameter exhibits a clear maximum for BST-07. Based on Slack's equation, these results are  directly related to the optimal ZT values for this composition. Our study provides a solid  framework for connecting composition,  lattice dynamics, and thermal-transport behaviour, towards establishing the microscopic origin of thermoelectric optimization in BST compounds.    
\end{abstract}


\maketitle

\section{INTRODUCTION}
Thermoelectric (TE) materials, stand out  as one of the most promising  energy-conversion materials, aiming to  meet  pressing global energy needs. The thermoelectric performance of a TE  material is governed by the coupled transport properties of the charge carriers and phonons, as described by the dimensionless figure of merit ZT (ZT=$\sigma$S$^2$T/($\kappa_e$ +$\kappa_l$), where S is the Seebeck coefficient, $\sigma$ is the electrical conductivity, T is the absolute temperature and $\kappa_e$ and $\kappa_l$ are the electronic and lattice thermal conductivities, respectively ). Thus, achieving a high ZT value  requires not only favorable electronic transport properties but also effective suppression of the lattice-mediated thermal conductivity \cite{He2017,Shi2025}. Various approaches have been  pursued to improve the  performance ($i.e.$, the ZT value) of TE materials, such as doping and alloying, novel synthesis methods etc. \cite{Lu2025}. Alloying provides a particularly powerful route to simultaneously and advantageously tune both  the electronic transport properties and the lattice thermal conductivity.  In addition to modifying the carrier concentration and the electronic structure, alloying alters the effective atomic masses, the interatomic force constants and lattice anharmonicity, all of which can influence phonon transport properties \cite{Xiong2024}. Therefore, understanding how alloying reshapes the lattice vibrational properties is essential for establishing the microscopic origin of thermoelectric optimization. 

Among near-room-temperature TE materials, Bi$_2$Te$_3$ is considered  an archetypal   material and it is the most widely used commercially.  Alloying Bi$_2$Te$_3$, specifically with Sb, is  one  possible route to achieve higher ZT for the next-generation  superior TE materials \cite{Zhou2025, Champness1965}. Therefore, the Bi-Sb-Te (BST) system provides an ideal platform for examining such composition-dependent lattice effects. Both Bi$_2$Te$_3$ and Sb$_2$Te$_3$ crystallize in the same rhombohedral $R\overline{3}m$ structure under  ambient conditions, consisting of Te1-Bi/Sb-Te2-Bi/Sb-Te1 quintuple layers separated by weak van der Waals gaps, and they form an extended substitutional Bi/Sb solid solution \cite{Mansour2014}, that retains the $R\overline{3}m$ structure \cite{Wei2025}. Optimized  Bi/Sb ratios produce some of the most widely used p-type thermoelectric materials near room temperature, with compositions around (Bi$_{0.3}$Sb$_{0.7}$)$_2$Te$_3$ and neighbouring BST alloys repeatedly serving as benchmark materials for high thermoelectric performance. The superiority of this composition range has commonly been discussed in terms of the carrier concentration, band structure, microstructure, and optimization of the electrical transport properties \cite{Poudel2008,Kim2015,Yu2017}. However, changing the Bi/Sb ratio unavoidably modifies the  phononic dispersion relation as well. Although Bi and Sb are isovalent, they substantially  differ  in atomic mass and local bonding characteristics. Thus,  alloying provides a continuous perturbation of both the inertial and the restoring-force terms governing the lattice vibrations. It is therefore natural to ask whether the favorable composition range is also associated with a distinctive lattice-dynamical response. 

In this context, the phononic properties  of BST compounds cannot be understood as a simple interpolation between the two end compounds based solely on changes in atomic mass. Element-resolved measurements and lattice-dynamical analyses of Bi$_2$Te$_3$ and Sb$_2$Te$_3$ have demonstrated substantial differences in their vibrational spectra. While the overall softening of the phonon density of  states (PDOS) of Bi$_2$Te$_3$ relative to Sb$_2$Te$_3$  is partially consistent with their mass difference, the accurate reproductions of  the optical phonons spectrum requires a significant change in the effective Bi-Te and Sb-Te force constants \cite{Fang2019, Bessas2012}. This highlights a competition between mass and bonding effects during Bi/Sb substitution. Therefore, Raman spectroscopy appears advantageous for probing these changes, given that Raman shifts contain intertwined contributions from mass substitution, changes in bond stiffness, local strain, and atomic disorder.   In the specific case of BST compounds, Raman spectroscopy is particularly sensitive to this competition, because the Raman-active modes ($A_{1g}^1, A_{1g}^2, E_{g}^1, E_{g}^2$) probe characteristic out-of-plane and in-plane atomic motions within the layered structure. Composition-dependent Raman investigations of BST at ambient conditions  have indeed revealed systematic phonon-frequency shifts, mode-dependent evolution, and disorder-related broadening throughout the solid solution \cite{Elyamny2022,Richter1977,Zhang2015}. 

Going one step further, the isothermal Grüneisen $\gamma$$_T$ parameter quantifies how a specific  phonon mode responds to changes in crystal volume. Whereas  Raman frequency at ambient conditions describes the curvature of the vibrational potential in  the equilibrium structure, $\gamma$$_T$ quantifies  how this curvature evolves when the interatomic distances are modified. It therefore provides an important measure of mode-dependent anharmonic lattice response. High-pressure Raman measurements and first-principles calculations have previously yielded mode-specific and macroscopic Grüneisen parameters for Bi$_2$Te$_3$, confirming that its lattice dynamics is highly sensitive  to compression \cite{Kullmann1984,Pradhan2012}. Likewise, studies of Sb$_2$Te$_3$-based thermoelectric systems have  demonstrated that doping can alter phonon anharmonicity and its connection to thermal transport \cite{Das2017}. Although  the Grüneisen parameters should not be interpreted as a direct measurement of lattice thermal conductivity they provide a key experimentally accessible evaluation  of the anharmonic lattice response that participates in phonon-phonon interactions. This correlation is particularly intriguing for high-performance Bi-Sb-Te compositions. 

Recent investigations on (Bi$_{0.25}$Sb$_{0.75}$)$_2$Te$_3$ have revealed strong coupling between low-lying optical and acoustic phonons and shown  that the low lattice thermal conductivity is associated with reduced phonon group velocities, along with enlarged scattering phase space, scattering rates, and Grüneisen parameters \cite{Bai2023}. Such observations suggest that the favourable thermoelectric properties of BST may involve not only optimized electronic transport and extrinsic phonon scattering from defects or microstructures but also intrinsic features of its anharmonic phonon spectrum. However,  results on an individual high-performance composition cannot establish whether this behavior is generic to the entire BST series or whether the anharmonic response itself exhibits a composition-selective enhancement. Despite extensive studies of composition-dependent Raman frequencies \cite{Richter1977, Zhang2015}, a systematic relationship between the  Bi/Sb ratio, Raman-active phonons, force constants, mode-resolved Grüneisen parameters, and the thermoelectric-relevant composition range remains insufficiently established.

In this study, we systematically investigate the composition-dependent lattice dynamics of (Bi$_{1-x}$Sb$_x$)$_2$Te$_3$ (BST-x) compounds with varying  Sb concentration  using Raman spectroscopy under  ambient and high-pressure conditions. The relative force constants of the  Raman-active modes for the compositions x=0 (Bi$_2$Te$_3$), 0.2, 0.5,  0.7, 0.8, 0.9 and 1(Sb$_2$Te$_3$)  are calculated based on a harmonic approximation. The results reveal that while the force constants of the  lower-frequency Raman modes ($E_g^1$ and $A_{1g}^1$) remain almost composition-independent, in agreement with previous Phonon Density of States (PDOS) studies \cite{Bessas2012}, a diametric softening ($E_g^2$) and  stiffening ($A_{1g}^2$) were determined for the higher-frequency ones. Moreover, the isothermal  Grüneisen parameters for the $A_{1g}^1$, $E_g^2$ and  $A_{1g}^2$ Raman modes were determined, by combining the Raman measurements under pressure in this study with the   Equation of State (EOS) for the ambient-conditions $R\overline{3}m$ phases reported in previous studies. Unlike a simple monotonic interpolation between the Bi$_2$Te$_3$ and Sb$_2$Te$_3$ end members, the Grüneisen parameters exhibit a pronounced composition dependence, with the strongest anharmonic response occurring near (Bi$_{0.3}$Sb$_{0.7}$)$_2$Te$_3$. Notably, this composition lies near the established high-performance p-type BST composition regime. This correspondence does not imply that the Grüneisen parameter alone determines the thermoelectric figure of merit; rather, it identifies an intrinsic lattice-dynamical feature that coincides  with a favorable composition range. The results therefore suggest that enhanced phonon anharmonicity may constitute an additional, previously underemphasized phononic factor in optimizing  BST thermoelectrics and provide a framework for connecting composition, bonding, lattice dynamics, and thermal-transport behavior.

\section{EXPERIMENTAL METHODS}
\subsection{Raman spectroscopy}
Raman studies were performed using a custom-made confocal micro-Raman system with the 660 nm line of a solid-state laser for excitation in the back-scattering geometry. The laser probing spot dimension was $\approx$ 4 $\mu$m. Raman spectra were recorded with a spectral resolution of 2 cm$^{-1}$ using a single-stage grating spectrograph equipped with a charge-coupled device (CCD) array detector. We kept the  laser power on specimens  below 0.2 mW to avoid any laser-induced decomposition. Ultra-low-frequency solid-state notch filters allowed us to measure Raman spectra down to 10 cm$^{-1}$ \cite{Hinton2019}. For both the ambient and high-pressure studies, the BST-x powder specimens were from the same batch used in our recent high-pressure x-ray diffraction study \cite{Wei2025}.

\subsection{High-pressure studies} 
Symmetric diamond anvil cells (DACs) with diamond culets of 300–400 $\mu$m in diameter were used. Between the two diamonds, a high-pressure chamber was constructed using a pre-indented Rhenium gasket with a thickness of 30–40 $\mu$m and a central hole diameter of $\approx$ 100$\mu$m. Ruby powder was also loaded inside the sample chamber and the shift of the ruby  line \cite{Syassen2008} was used to determine the pressure \textit{in-situ}. The BST-x powders with composition x = 0.2, 0.7 and  0.9 were individually loaded in the center of the chamber. The remaining volume was filled with neon (Ne), precompressed to $\approx$ 2kbar (using a gas loader), acting as the pressure-transmitting medium (PTM).  Neon (Ne) remains fairly hydrostatic up to $>$20 GPa \cite{Klotz2009}, i.e., above the maximum pressure in this study.

\section{RESULTS AND DISCUSSION}

\subsection{Raman spectroscopy of BST under ambient conditions }
Figure 1(a) shows the Raman spectra of BST-x compounds (with x=0 (Bi$_2$Te$_3$), 0.2, 0.5, 0.7, 0.8,  0.9 and 1(Sb$_2$Te$_3$)) at ambient conditions. All BST compounds in this study adopt the rhombohedral $R\overline{3}m$ structure (tetradymite-type) throughout the solid-solution series. According to  group theory, for the  $R\overline{3}m$ structure, 4 zone-center Raman-active modes are expected with symmetries $\widetilde{\Gamma}_{R\overline{3}m}$= $2A_{1g}$$+2E_g$.  The  $E_g$ and the $A_{1g}$ Raman-active modes correspond predominantly to in-plane and out-of-plane atomic displacements, respectively, see Fig. 1 bottom \cite{Keresztury2001, Richter1977,Abdullaev2022}.  All four characteristic modes are resolved in the recorded spectra and can be assigned to the symmetries (in increasing frequency order)   $E_g^1$, $A_{1g}^1$, $E_g^2$ and $A_{1g}^2$. The persistence of the same symmetry Raman-active modes  across the investigated compositions is consistent with the preservation of the basic tetradymite framework upon Bi/Sb substitution. 

\begin{figure}[ht]
\centering
\includegraphics[width=0.8\linewidth]{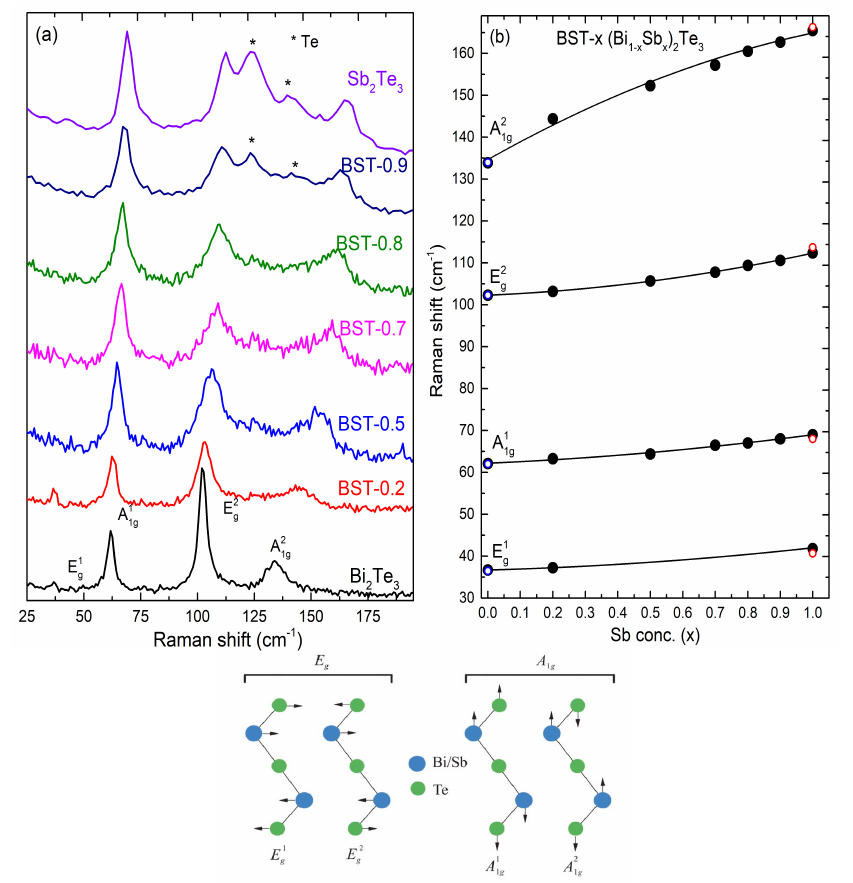}
\caption{(a) Raman spectra of BST-x for x=0 (Bi$_2$Te$_3$), 0.2,0.5,  0.7, 0.8, 0.9 and 1(Sb$_2$Te$_3$)) at ambient conditions. The asterisks denote few impurity peaks from Te. (b) Frequency as a function of Sb concentration (x) of the four Raman-active modes. The open symbols correspond to previously reported frequencies for Bi$_2$Te$_3$ (blue) \cite{Kullmann1984} and Sb$_2$Te$_3$ (red) \cite{Norimatsu2015}.  The schematic at the bottom shows the Raman-active modes eigenvectors, adapted from Ref. \cite{Abdullaev2022}. }
\end{figure}

The progressive blue shift of all four Raman modes  indicates an overall increase of the optical phonons frequency   as  Sb concentration increases. This behavior can be  understood primarily in terms of the reduced effective mass associated with substituting heavier Bi atoms with lighter Sb atoms, consistent with previous Raman and lattice-dynamical investigations of the Bi$_2$Te$_3$–Sb$_2$Te$_3$ solid-solution system \cite{Richter1977, Urkude2023}. However, previous studies strongly indicate that mass substitution alone does not fully account for changes  in  the PDOS \cite{Bessas2012}. Specifically, the $A_{1g}^2$ mode shifts from approximately 134 in Bi$_2$Te$_3$ to 165 cm$^{-1}$  in Sb$_2$Te$_3$, while the other three modes show a weaker composition dependence. This different composition-dependent behavior shall be related to the force constant of each mode.

\subsection{Force constant evaluation } 
To further understand this  mode-dependent behavior, we estimated the force constant for each mode using a harmonic approximation.  All of the eigenvectors of the Raman active modes in the tetradymite-type structure  involve vibrations of two pairs of Bi/Sb-Te1 atoms, either perpendicular ($E_g$) or parallel ($A_{1g}$) to the c-axis, see Fig. 1.  We used a  harmonic approximation $\omega_i$ = $\sqrt{\frac{k_i}{M_i}}$, where $k_i$ and $M_i$ denote the force constant and displacement-weighted effective mass associated with mode i, to estimate the force constant $k_i$. The effective mass is then  calculated according to the following equation: $M_i$ = $\frac{\sum_{j} m_j u_{j,i}^{2}}{\sum_{j} u_{j,i}^{2}}$, where the $u_{j,i}$ defines the atomic displacement vector of atom j in mode i. The  cation can be either Bi or Sb and the anion is always Te, thus we used the Bi/Sb ratio (e.g.,  0.3*m$_\alpha$Bi+0.7*m$_\alpha$Sb  for BST=0.7)  to calculate the mass of the cation. From the above, the force constant can be obtained by using the Raman frequency and the effective mass. We used equal displacement vector approximation for the case of the  $E_g^1$ and $E_g^2$ modes \cite{Sosso2009}.  For $A_{1g}^1$ and  $A_{1g}^2$, the atomic displacement vectors for Sb$_2$Te$_3$ were used to approximate the whole concentration series \cite{Huang2023}. The results of the  normalized mode force constants, relative to Bi$_2$Te$_3$,  are shown in Figure 2

\begin{figure}[ht]
\centering
\includegraphics[width=0.7\linewidth]{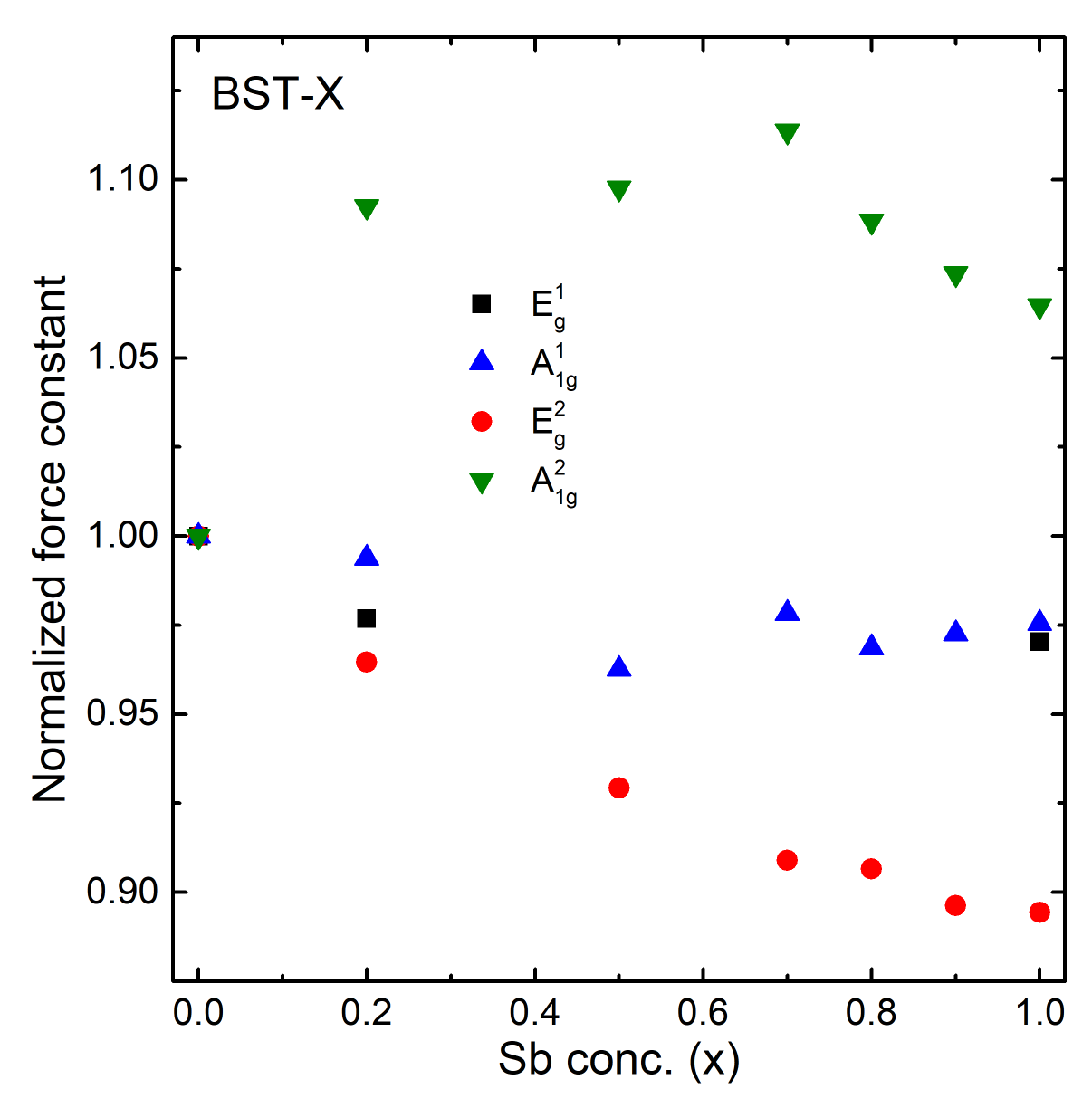}
\caption{Normalized, to Bi$_2$Te$_3$, force constants of the 4 Raman active modes of BST-x compounds as a function of Sb concentration.    }
\end{figure}

The force constants of the $E_g^1$ and the $A_{1g}^1$ modes, remain practically constant, taking into account the approximate nature of the calculations,  throughout the compositional range. The force constant of the $E_g^2$ mode strongly and monotonically decreases by $\sim$10\% with increasing Sb concentration between  x=0 to 1.  In contrast,  the higher-frequency $A_{1g}^2$ mode shows a distinctly different non-monotonic evolution, with an initial  sharp increase, reaching a maximum near x=0.7, followed by a slight decrease.  Within the harmonic approximation model, this indicates that for the $E_g^1$ and the $A_{1g}^1$ modes, the frequency increase  directly reflects  the reduction in effective mass. On the other hand, for the $E_g^2$ mode the reduction of the effective mass  outweighs the decrease in the force constant.  Surprisingly,  the frequency increase of the $A_{1g}^2$ mode is sufficiently pronounced, due to an enhancement of its force constant, in addition to the decrease of the effective mass,  over much of the composition range. This is further supported by the superlinear trend of the mode frequency as a function of Sb concentration for the $E_g^1$, $E_g^2$ and $A_{1g}^1$ modes, as opposed to the sublinear for the $A_{1g}^2$ mode. Notably, despite these contrasting force constant trends, all four modes undergo phonon blue-shifting  with increasing Sb concentration

As discussed in Ref. \cite{Bessas2012}, the PDOS for Sb$_2$Te$_3$ below $\sim$ 100 cm$^{-1}$ could be reproduced by  the scaling of the Bi$_2$Te$_3$ PDOS, using a direct homology relation for the differences in effective masses. This agrees  with our results throughout the compositional range. On the other hand, in the same study  a $\sim$ 20\% softening of the Sb  force constant,  compared to Bi, was  needed to explain the PDOS shift  between Bi$_2$Te$_3$ and Sb$_2$Te$_3$, for  phonon energies above $\sim$ 100 cm$^{-1}$,  even after accounting for  the slight difference in lattice parameters.  This well explains the progressive force constant  softening  behavior of the $E_g^2$  mode with increasing  Sb concentration. However, the force constant of $A_{1g}^2$ shows exactly the opposite trend, namely a significant increase up to x = 0.7. These contradicting  force-constant trends may also reflect composition-dependent modifications of the local bonding environment and lattice structure that will result in changes of the phonon dispersion relations, especially at the center of the Brillouin zone, beyond the level of the collective PDOS. 

Previous neutron diffraction investigations of BST compounds revealed a systematic contraction of the intralayer Bi/Sb-Te bond lengths upon Sb substitution, accompanied by changes in the inter-quintuple Te-Te separation \cite{SerranoSanchez2017}. Such structural modifications may influence the vibrational restoring forces by changing  the intra- and inter-quintuple interactions. Importantly, these structural effects are expected to influence different vibrational modes unequally. This  was also demonstrated in a recent study, which reported that the low-frequency $A_{1g}^1$ mode in Sb$_2$Te$_3$ is particularly sensitive to inter-quintuple decoupling, while the high-frequency $A_{1g}^2$ mode responds more strongly to intralayer bonding distortion \cite{Mertens2024}. Accordingly, the distinct non-monotonic behavior of the $A_{1g}^2$ force constant may reflect  its different sensitivity to the local structural and bonding modifications induced by Sb substitution.

Nevertheless, since the present force constants were calculated using composition-independent atomic displacements, possible changes in the phonon displacement eigenvectors upon alloying are not explicitly considered. Therefore, the calculated trends should be interpreted as model-dependent variations in the estimated vibrational stiffness rather than direct evidence of individual bond strengthening or weakening. However, the fact that our simplified model accurately reproduces an almost concentration-independent  force constant for the low frequency  $E_g^1$ and $A_{1g}^1$ modes (both below 100 cm$^{-1}$) in full agreement with the comparative study of the     PDOS in Ref. \cite{Bessas2012}, further strengthens the qualitative validity of our approach. 

\subsection{Raman spectroscopy of BST under pressure } 
To further investigate the lattice vibrational response to compression, we preformed high-pressure Raman measurements  on BST compounds with x=0.2, 0.7, and 0.9. From the four expected Raman-active modes, three ($A_{1g}^1$, $E_g^2$ and $A_{1g}^2$) were systematically observed under pressure and will be used in the following analysis and discussion. The pressure ranges up to $\sim$ 7-8 GPa, was constrained to the maximum  pressure at which the ambient conditions  $R\overline{3}m$ structure  is retained without  coexistence of the $C2/m$ phase \cite{Wei2025}. Figures 3 and 4(a) show the corresponding Raman spectra collected at selected pressures. All observed Raman-active modes for all BST compounds examined in this study   shift monotonically toward higher frequencies upon compression, indicating a pressure-induced blue-shift, consistent  with previous Raman  studies under pressure for Bi$_2$Te$_3$ and Sb$_2$Te$_3$. This behavior is consistent with changes (increases)  in the interatomic restoring forces resulting from reduced  interatomic distances under pressure.

\begin{figure}[ht]
\centering
\includegraphics[width=0.9\linewidth]{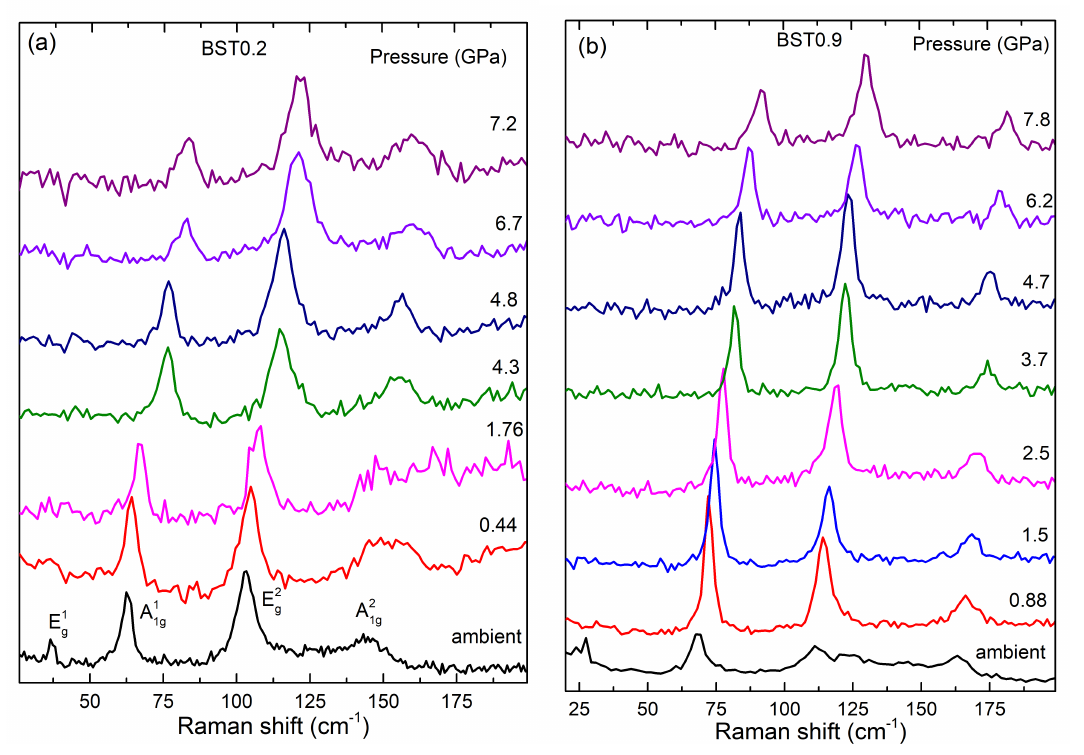}
\caption{ Raman spectra of (a) BST-0.2 and (b) BST-0.9 at selected pressures upon pressure increase.  }
\end{figure}

\begin{figure}[ht]
\centering
\includegraphics[width=0.9\linewidth]{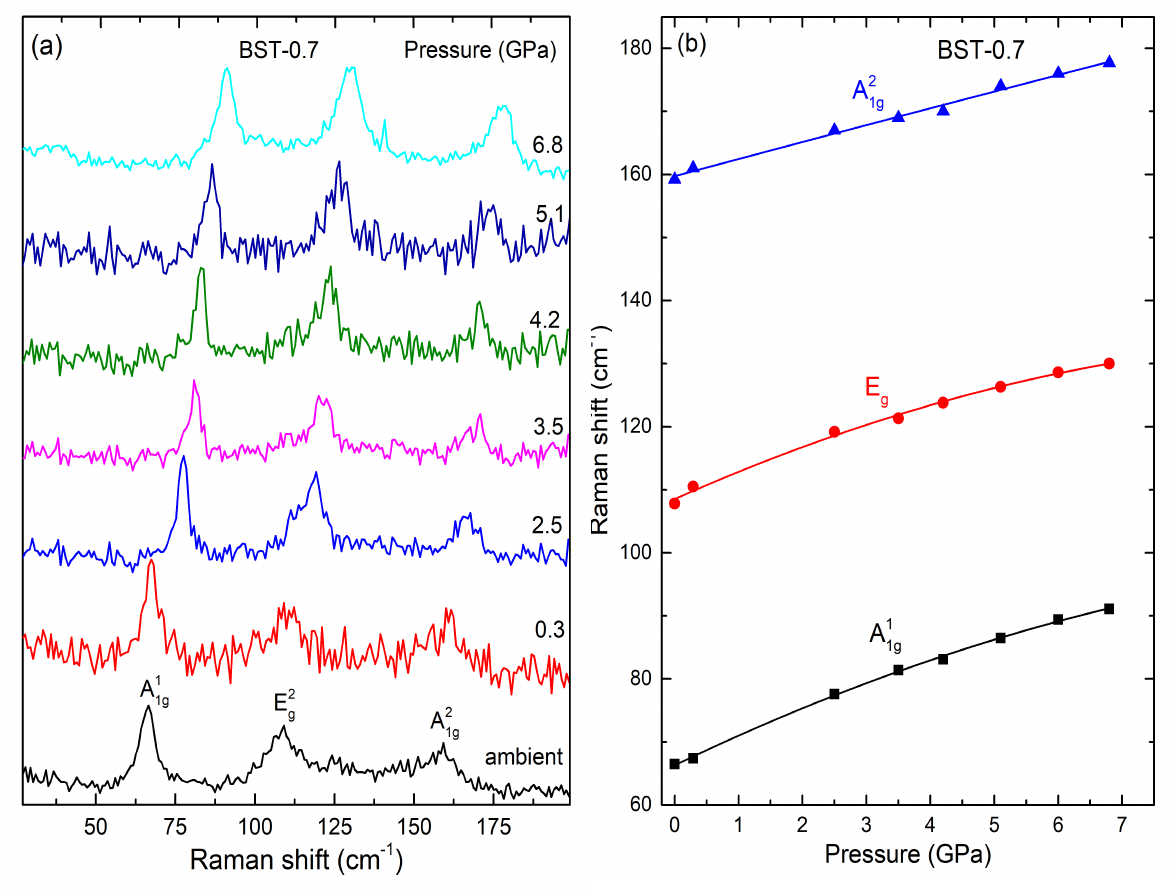}
\caption{(a) Raman spectra of BST-0.7 at selected pressures upon pressure increase. (b) Raman mode frequencies of BST-0.7 as a function of pressure.  The solid curves are fits using a second-order polynomial; see text for details.   }
\end{figure}

To quantitatively describe   the pressure-induced frequency shifts, the pressure dependence of the frequency of each resolved Raman mode was fitted using a second-order polynomial: $\omega(P)$ = $\omega_0$ + $\alpha$ P + $\beta$ $P^2$, where $\omega_0$ is the ambient-pressure frequency and $\alpha$ and $\beta$ are the corresponding first- and second-order pressure coefficients  \cite{Souza2012}. Figure 4(b) shows the corresponding frequency as a function of pressure  for BST-0.7 and Table I summarizes   the  results for all BST-x compounds  studied under pressure,  along  with the results for Bi$_2$Te$_3$ and Sb$_2$Te$_3$ from previous studies \cite{Vilaplana2011, Souza2012}.

\begin{table*}[ht]
\centering
\normalsize
\caption{Bulk modulus B$_0$ (GPa), ambient-pressure Raman mode frequency $\omega$$_0$ (cm$^{-1})$, frequency-pressure slope $\partial$$\omega$/$\partial$P (cm$^{-1}$/GPa) and  Grüneisen parameter $\gamma_{iT}$ for the BST-X compounds discussed in the text. The values for Bi$_2$Te$_3$ are from Refs. \cite{Einaga2011} (B$_0$) and \cite{Vilaplana2011} (Raman measurements) and for Sb$_2$Te$_3$ from Refs. \cite{Sakai1981} (B$_0$) and \cite{Souza2012} (Raman measurements).  }
\medskip
\begin{tabular}{llllll}
Composition& Bi$_2$Te$_3$&BST-0.2&BST-0.7&BST-0.9&Sb$_2$Te$_3$ \\\hline
B$_0$& 45.0& 45.9& 43.6 & 42.8 & 40\\ 
$\omega$$_0$ $A_{1g}^1$&61.3&63.3&66.5&68&70.6\\ 
$\omega$$_0$ $E_{g}^2$&101.1&103.2&107.8&110.6&114.2\\ 
$\omega$$_0$ $A_{1g}^2$& 132.1&144.1&159.2&162.7&166.4\\ 
$\partial\omega/\partial$P $A_{1g}^1$&3.71&3.57&4.87&4.29&4.42\\ 
$\partial\omega/\partial$P $E_{g}^2$ &3.49&3.34&4.52&3.80&2.86\\ 
$\partial\omega/\partial$P $A_{1g}^2$&2.92&3.32&2.72&3.73&	2.75\\ 
$\gamma_{T}$ $A_{1g}^1$ & 2.72&2.59&3.19&2.70&2.50\\ 
$\gamma_{T}$ $E_{g}^2$ & 1.55&1.48&1.83&1.47&1.00\\ 
$\gamma_{T}$ $A_{1g}^2$ & 0.99&1.06&0.74&0.98&0.66\\ \hline
\end{tabular}
\end{table*}

\subsection{Grüneisen parameters} 
Using $\gamma_{T}$ removes part of the trivial difference associated with the absolute phonon frequency and instead quantifies how strongly each vibrational mode responds to a fractional change in volume. Pressure-dependent Raman studies of Bi$_2$Te$_3$ have previously demonstrated substantial mode-dependent frequency shifts and yielded mode Grüneisen parameters for the rhombohedral phase \cite{Vilaplana2011}. First-principles lattice-dynamical calculations have likewise identified appreciable vibrational anharmonicity in Bi$_2$Te$_3$, giving a macroscopic average Grüneisen parameter of approximately 1.56 \cite{Pradhan2012, Urkude2023}. These studies establish that the tetradymite lattice is intrinsically sensitive to changes in interatomic spacing, but they do not address how this response evolves continuously across the (Bi$_{1-x}$Sb$_x$)$_2$Te$_3$ solid solution.

Based on the results in the previous section, the frequency response to pressure differs substantially among the Raman modes and, more importantly, varies with Bi/Sb composition. The corresponding Raman modes isothermal  Grüneisen parameters for all observed Raman modes of the $R\overline{3}m$ phase  were calculated according to  $\gamma_{iT}$ = -( $\frac{\partial \ln \omega_i}{\partial \ln V} $) $_T$ = $\frac{B_0}{\omega_{i,0}} ( \frac{\partial \omega_i}{\partial P} )_T$, where $B_0$ is the ambient-pressure bulk modulus. For the compositions x=0.2, 0.7 and  0.9 we  determined the $B_0$ from the Equation of State (EOS) results  for the $R\overline{3}m$ phases reported in  our previous high-pressure XRD study \cite{Wei2025}. For the two end members (Bi$_2$Te$_3$ and Sb$_2$Te$_3$) we have used the EOSs from previous studies \cite{Einaga2011, Sakai1981} for the $R\overline{3}m$ phases.  We note that when a 3rd-order Birch–Murnaghan EOS \cite{Birch1947}  is used to fit the P-V data, substantial differences in B$_0$ values might arise between various studies \cite{Vilaplana2011}, due to the well-known interplay (opposite correlation) between B$_0$ and B$_0$’ values \cite{Zou2026}. This variation, will  have a profound effect on the determined  $\gamma_{iT}$ values for different BST compounds and Raman modes.  For this reason, a 2nd-order Birch–Murnaghan EOS (B'=4, fixed) was fitted to all EOS data used in this study \cite{Wei2025, Einaga2011, Sakai1981}. Although a  2nd-order EOS might not produce the optimum least-squares fit, it minimizes the effect of the variation in the B$_0$ values determined from different experimental studies. Nevertheless, our aim is a comparative study of  $\gamma_{T}$ values and not an absolute determination. Using this methodology, the determined  $\gamma_{T}$ values for all observed Raman modes under pressure,   are listed in Table I and shown in Fig. 5.

\begin{figure} [ht]
\centering
\includegraphics[width=0.7\linewidth]{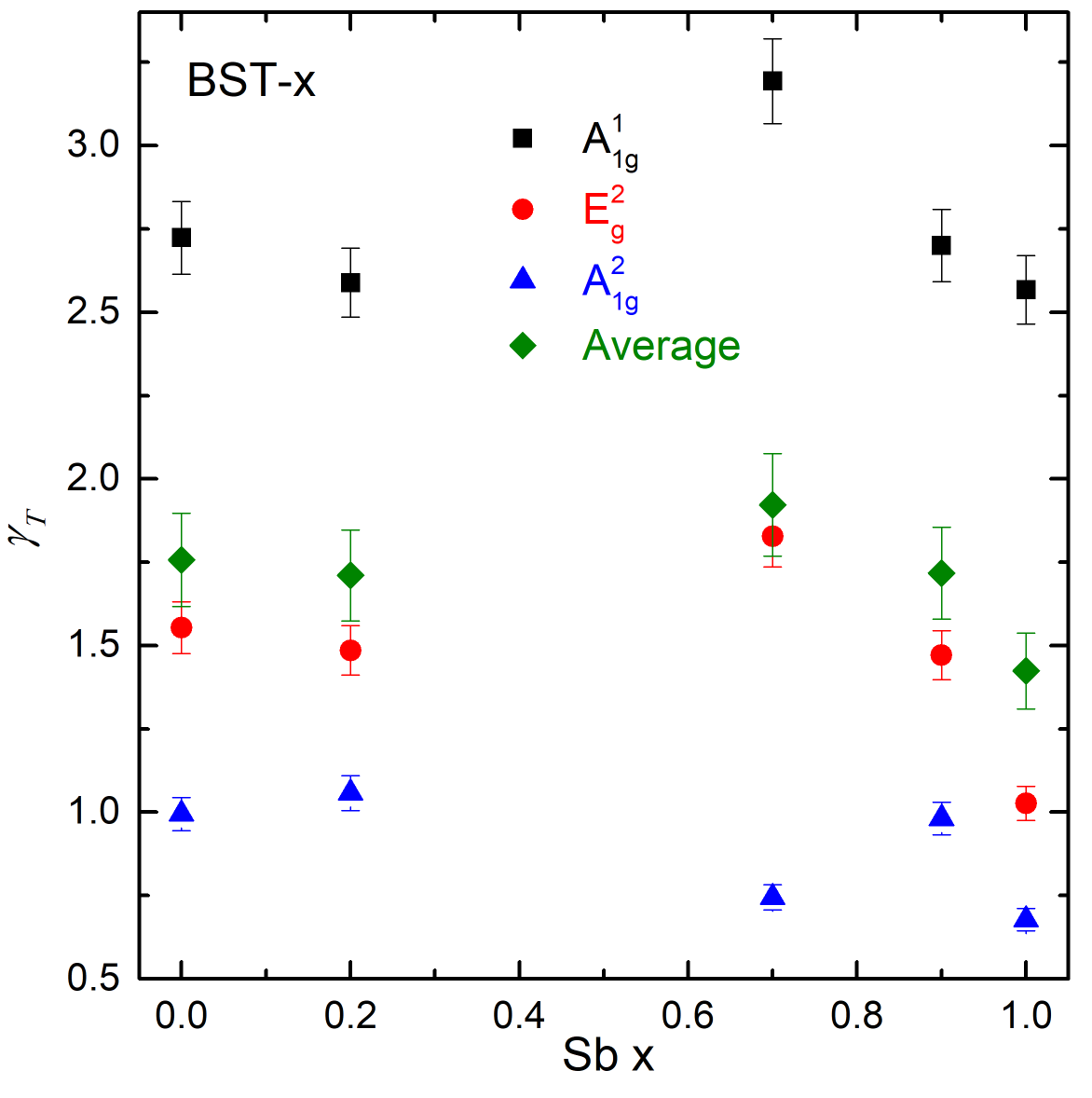}
\caption{(a) $\gamma_{T}$ values, as a function of Sb concentration, for the  $A_{1g}^1$, $E_g^2$ and $A_{1g}^2$ Raman modes, see text for details, together with the average one from these modes.     }
\end{figure}

From Fig. 5, after taking into account relevant uncertainties and experimental errors, certain conclusions could be made. The $\gamma_{T}$ of both the $A_{1g}^1$ and  $E_g^2$ Raman modes shows a maximum at the composition x=0.7, while the one of the $A_{1g}^2$ appears practically composition independent. The maximum at x=0.7 is also reflected in the average $\gamma_{T}$ for the three modes. Although a definite explanation for the composition independence  of  $\gamma_{T}$ for the  $A_{1g}^2$ mode is not possible; it is plausible to assume that this is related to the stiffening  of the force constant upon increase of the Sb content. Based on  Slack's equation \cite{Slack1973,Skoug2010}, the thermal conductivity is inversely proportional to $\gamma_{T}$$^2$. Although thermal conductivity is mainly correlated  with acoustic phonons,  our analysis provides a clear indication about the origin of the lower thermal conductivity for BST-07, which, as mentioned, is highlighted as the one with the maximum ZT values. Further studies are needed, mainly on the $\gamma_{T}$ of acoustic phonons,  to fully elucidate this correspondence.

\section{Conclusion} 
Based on the force-constant analysis using the  Raman measurements at ambient conditions for the BST compounds in this study,  a diverging  response of the force constant as a function of the composition was concluded, beyond the PDOS scaling proposed in previous studies. This indicates that the force constants may reflect the distinct sensitivity of each Raman mode to  local structural and bonding modifications induced by Sb substitution. The average Grüneisen parameter, calculated from the corresponding $\gamma_{T}$ of the $A_{1g}^1$,  $E_g^2$ and $A_{1g}^2$ Raman modes, exhibits a clear maximum for BST-07. Based on Slack's equation, this is directly related to the optimal ZT values for this composition. Our study represents a critical step forward towards establishing the microscopic origin of thermoelectric optimization in BST compounds. Moreover, it provides the framework for connecting composition, bonding, lattice dynamics, and thermal-transport behavior.

\section*{Acknowledgements}
We thank Azkar Saeed Ahmad for his help with specimens  loadings  and Muhamed Dawod and  Yaron Amouyal for providing the BST specimens. The work  was supported by funding from the Guangdong Technion Israel Institute of Technology, the Guangdong Provincial Key Laboratory of Materials and Technologies for Energy Conversion, MATEC (Grant No. 2022B1212010007, Guangdong Department of Science and Technology) and the Technion-GTIIT seed grant program.



\subsection*{Availability of data and materials}
The data that support the findings of this study are available from the corresponding author upon reasonable request.

\subsection*{AI and AI-assisted tools Statement} 
AI-assisted tools were used solely for language and grammar editing. 


\subsection*{Conflicts of interest}   
The authors have no conflicts to disclose.



\bibliography{BSTRaman}

\end{document}